\documentstyle[12pt]{article}
\def\be{\begin{equation}}
\def\ee{\end{equation}}
\def\ba{\begin{array}}
\def\ea{\end{array}}

\begin{document}
\title{Reconciling $G_A/G_V, ~<r_n^2>$ and $\mu_{p,n}$ in $\chi$QM with
One Gluon Generated Configuration Mixing.}
\author{Harleen Dahiya  and Manmohan Gupta    \\
{\it Department of Physics,} \\
{\it Centre of Advanced Study in Physics,} \\
{\it Panjab University,Chandigarh-160 014, India.}}   
 \maketitle

\begin{abstract}
The spin polarization functions $(\Delta u, ~\Delta d, ~\Delta s)$
for proton are calculated in the chiral quark model ($\chi$QM)
with SU(3) symmetry breaking as well as  
configuration mixing generated by one gluon exchange forces for the NMC
and the most recent E866 data.
 Besides reproducing the spin polarization functions $\Delta u, ~\Delta d, 
~\Delta s$ as well as $G_A/G_V$, it can accomodate nucleon magnetic moments
 and neutron charge radius as well,
 thus resolving the compatibility problem of these parameters
 which could not be achieved in constituent quark models.
\end{abstract}

 The Constituent Quark Model (CQM), despite its
 impressive performance in explaining low
 energy hadronic matrix elements, is unable to account for the
 EMC effect or the ``proton spin crisis'' {\cite{EMC}} which
 indicated that only a small portion of the proton spin is
 carried by the valence quarks, as well as the presence
 of significant negative strange quark polarization in the
 proton quark sea  {\cite{sea}}. CQM is also
 unable to explain the results of the NMC {\cite{NMC}} and E866
 {\cite{E866}} experiments which have shown that the  Gottfried sum rule
{\cite{GSR}} is violated, indicating that the $\bar d$
density is larger than $\bar u$ density in the nucleon sea.
Apart from the problems faced by CQM in explaining the spin
content of nucleons, it is also saddled with another problem,
$viz.$, it has been shown {\cite{mgupta2}}
that it is not possible to have a simultaneous reconciliation
of $G_A/G_V$, charge form factors of nucleon  and
magnetic moments of nucleons ($\mu_{p,n}$) in the CQM. This problem
becomes more acute if one considers
neutron charge radius $(<r^2_{n}>)$.

The EMC effect  and related issues have been
successfully addressed in  the chiral quark model
($\chi$QM) {\cite{{cheng},{quigg},{johan},{song}}},  originally
conceived by Weinberg {\cite{wein}}, subsequently developed by
Manohar and Georgi {\cite{manohar}}. In particular, the $\chi$QM
 is able to explain not only the  spin content of the
nucleon but is also able to give a fair description of flavor
content of the nucleons, violation of Gottfried sum rule, strange
 quark content in the nucleon, $G_A/G_V$ as well as the magnetic
 moments of the baryons etc.
 {\cite{{cheng},{quigg},{johan},{song}}.

Recently it has been claimed that, within the $\chi$QM,
neutron charge radius can also be reproduced through
the Foldy term {\cite{bernard}}, however very recently it has
been shown by Isgur   {\cite{Isgur}} that the Foldy term gets
cancelled against a contribution to the Dirac form factor
$F_1$ to leave intact the interpretation of neutron electric
charge form factor $G_E^n$ as arising
from the neutron's rest frame charge  distribution.  Therefore,
in the light of the work of Isgur, it becomes essential that
$<r_n^2>$ should be reproduced by the form factor $F_1$. Thus,
even in the  $\chi$QM, the question of compatibility of
$G_A/G_V$, ~$<r^2_n>$ and $\mu_{p,n}$ remains.

In the context of CQM, it is well known that there are several
low energy parameters which can be explained by including one gluon
mediated configuration mixing \cite{{DGG},{Isgur1},{yaouanc},{mgupta},
{mgupta1}}.
 It has also been shown that one gluon mediated configuration mixing
 is able to generate $<r^2_n>$ as well as it improves the fit of
 magnetic moment of baryons {\cite{{yaouanc},{mgupta},{mgupta1}}}.
In view of the fact that the $\chi$QM
incorporates the basic features of constituent quark model,
therefore it becomes interesting to examine the implications
of configuration mixing in $\chi$QM ($\chi$QM$_{gcm}$).
The  purpose of the present communication on the one hand is to
examine, within the $\chi$QM, the implications of one gluon mediated
configuration mixing for spin polarizations and related issues while on the
other hand it is to investigate the issue of compatibility of
$G_A/G_V$, neutron charge radius and $\mu_{p,n}$.

To understand the implications of one gluon mediated configuration mixing
 for $G_A/G_V$, we
first calculate the quark spin polarizations in 
the  $\chi$QM. The basic process, in the $\chi$QM, is the
emission of a Goldstone  Boson which further splits into
$q \bar q$ pair, for example,

\be
  q_{\pm} \rightarrow GB^{0}
  + q^{'}_{\mp} \rightarrow  (q \bar q^{'})
  +q_{\mp}^{'}.
\ee
  The above process can be expressed through the Lagrangian
\be
{\cal L} = g_8 \bar q \Phi q,
\ee
where $g_8$ is the coupling constant,

\be
 q =\left( \ba{c} u \\ d \\ s \ea \right), 
 \ee
and

\be
 \Phi = \left( \ba{ccc} \frac{\pi^0}{\sqrt 2}
+\beta\frac{\eta}{\sqrt 6}+\zeta\frac{\eta^{'}}{\sqrt 3} & \pi^+
  & \alpha K^+   \\
\pi^- & -\frac{\pi^0}{\sqrt 2} +\beta \frac{\eta}{\sqrt 6}
+\zeta\frac{\eta^{'}}{\sqrt 3}  &  \alpha K^0  \\
 \alpha K^-  &  \alpha \bar{K}^0  &  -\beta \frac{2\eta}{\sqrt 6}
 +\zeta\frac{\eta^{'}}{\sqrt 3} \ea \right).
 \ee

SU(3) symmetry breaking is introduced by considering
different quark masses $m_s > m_{u,d}$ as well as by considering
the masses of  non-degenerate Goldstone Bosons
 $M_{K,\eta} > M_{\pi}$ {\cite{{johan},{song},{cheng1}}} , whereas 
  the axial U(1) breaking is introduced by $M_{\eta^{'}} > M_{K,\eta}$
{\cite{{cheng},{johan},{song},{cheng1}}}.
The parameter a(=$|g_8|^2$) denotes the transition probability
of chiral fluctuation
or the splittings  $u(d) \rightarrow d(u) + \pi^{+(-)}$, whereas 
$\alpha^2 a$ denotes the probability of transition 
$u(d) \rightarrow s  + K^{-(0)}$.
Similarly $\beta^2 a$ and $\zeta^2 a$ denote the probability of
$u(d,s) \rightarrow u(d,s) + \eta$ and
$u(d,s) \rightarrow u(d,s) + \eta^{'}$ respectively.

The one gluon exchange forces {\cite{DGG}}
generate the mixing of the octet in $(56,0^+)_{N=0}$ with
the corresponding octets in $(56,0^+)_{N=2}$,
$(70,0^+)_{N=2}$ and  $(70,2^+)_{N=2}$
harmonic oscillator bands {\cite{Isgur1}}. The
corresponding wave function of the nucleon is given by
 
 \[|B>=(|56,0^+>_{N=0} cos \theta +|56,0^+>_{N=2} sin \theta)
 cos \phi \] 
\be
 +(|70,0^+>_{N=2} cos \theta +|70,2^+>_{N=2}  sin \theta)
 sin \phi.
 \ee
In the above equation it should be noted that
$(56,0^+)_{N=2}$ does not affect the
spin-isospin structure of  $(56,0^+)_{N=0}$, therefore the mixed
nucleon wave function can be expressed in terms of  $(56,0^+)_{N=0}$
and  $(70,0^+)_{N=2}$, which we term as non trivial mixing
{\cite{mgupta1}} and is given as

\begin{equation}
\left|8,{\frac{1}{2}}^+ \right> = cos \phi |56,0^+>_{N=0}
+ sin \phi|70,0^+>_{N=2},
\end{equation}
 where
 \be
 |56,0^+>_{N=0,2} = \frac{1}{\sqrt 2}(\chi^{'} \phi^{'} +
\chi^{''} \phi^{''}) \psi^{s},
\ee

\be
|70,0^+>_{N=2} =  \frac{1}{2}[(\psi^{''} \chi^{'} +\psi^{'}
\chi^{''})\phi^{'} + (\psi^{'} \chi^{'} -\psi^{''} \chi^{''})
\phi^{''}].
\ee
 The spin and isospin wave functions, $\chi$ and $\phi$, are
given below
    \[\chi^{'} =  \frac{1}{\sqrt 2}(\uparrow \downarrow \uparrow
    -\downarrow \uparrow \uparrow),~~~  \chi^{''}
    =  \frac{1}{\sqrt 6} (2\uparrow \uparrow \downarrow
  -\uparrow \downarrow \uparrow
  -\downarrow \uparrow \uparrow), \] \\
\[\phi^{'}_p = \frac{1}{\sqrt 2}(udu-duu),~~~
\phi^{''}_p = \frac{1}{\sqrt 6}(2uud-udu-duu),\]
\[\phi^{'}_n = \frac{1}{\sqrt 2}(udd-dud),~~~
 \phi^{''}_n = \frac{1}{\sqrt 6}(udd+dud-2ddu).\]
 For the definition of the spatial part of the wave function,
 ($\psi^{s}, \psi^{'}, \psi^{''})$ as well as the
 definitions of the spatial overlap integrals we  refer the
 reader to references \cite{yaouanc} and \cite{mgupta1}.

 The contribution to the proton spin, defined through the equation
 \be
\Delta q =q_\uparrow - q_\downarrow  +\bar{q}_\uparrow
 - \bar{q}_\downarrow,
 \ee
using Equation(4) and following Linde $et al.$ {\cite{johan}}, can
be expressed as 

\be
   \Delta u ={cos}^2 \phi \left[\frac{4}{3}-\frac{a}{3}
   (7+4 \alpha^2+ \frac{4}{3} \beta^2
   + \frac{8}{3} \zeta^2)\right]
   + {sin}^2 \phi \left[\frac{2}{3}-\frac{a}{3} (5+2 \alpha^2+
  \frac{2}{3} \beta^2 + \frac{4}{3} \zeta^2)\right], 
\ee

\be
  \Delta d ={cos}^2 \phi \left[-\frac{1}{3}-\frac{a}{3} (2-\alpha^2-
  \frac{1}{3}\beta^2- \frac{2}{3} \zeta^2)\right]  + {sin}^2 \phi
  \left[\frac{1}{3}-\frac{a}{3} (4+\alpha^2+
  \frac{1}{3} \beta^2 + \frac{2}{3} \zeta^2)\right], 
\ee
and

\be
   \Delta s  = -a \alpha^2.
\ee
Before one presents and discusses the results pertaining to
Equations (10), (11) and (12), for a better appreciation of the role of 
configuration mixing and symmetry breaking we have considered
the case of SU(3) symmetry as well.
The SU(3) symmetric calculations can easily be obtained from the
equations (10), (11) and (12) by considering $\alpha, \beta =1$.
The corresponding equations can be expressed as
 \be
   \Delta u ={cos}^2 \phi \left[\frac{4}{3}
   -\frac{a}{9} (37+8 \zeta^2)\right]  + {sin}^2 \phi
  \left[\frac{2}{3}-\frac{a}{9} (23+4 \zeta^2)\right], 
\ee

\be
  \Delta d ={cos}^2 \phi \left[-\frac{1}{3}
  -\frac{2a}{9} (\zeta^2-1)\right]  + {sin}^2 \phi
  \left[\frac{1}{3}-\frac{a}{9} (16+2 \zeta^2)\right], 
\ee
and

\be
   \Delta s = -a.
\ee

\begin{table}
{\tiny
\begin{center}
\begin{tabular}{|c|c|c|c|c|c|c|c|c|c|c|c|c|}       \hline
 & & \multicolumn{5}{c|} {Without configuration mixing} &
\multicolumn{6}{c|} {With configuration mixing}\\ \cline{3-13} 
Para- & Expt & CQM & \multicolumn{2}{c|} {$\chi$QM}  &
\multicolumn{2}{c|} {$\chi$QM}
& $\phi$ & CQM$_{gcm}$ & \multicolumn{2}{c|} {$\chi$QM$_{gcm}$}     &
\multicolumn{2}{c|} {$\chi$QM$_{gcm}$} \\
 meter& value&   & \multicolumn{2}{c|} {with SU(3)}  &
\multicolumn{2}{c|} {with SU(3)} & & & \multicolumn{2}{c|} {with SU(3)}  &
\multicolumn{2}{c|} {with SU(3)} \\
 & &   & \multicolumn{2}{c|} {symmetry}  &
\multicolumn{2}{c|} {symmetry} & &
& \multicolumn{2}{c|} {symmetry}  &
\multicolumn{2}{c|} {symmetry} \\  
&&&\multicolumn{2}{c|}{}& \multicolumn{2}{c|} {breaking} &&&
\multicolumn{2}{c|}{}& \multicolumn{2}{c|} {breaking} \\ 
\hline
& & & NMC  & E866  & NMC &E866 & & & NMC & E866 & NMC & E866 \\
\cline{4-7} \cline{10-13}
 & & & & & & & 20$^0$ & 1.26 &
 .74 & .76 & .90, .86$^*$ & .92, .88$^*$  \\
 $\Delta$ u & 0.85  & 1.33 & .79 & .81 & .96 & .99 & 18$^0$ 
& 1.27 &
.75 &.77 & .91, .87$^*$ & .93, .89$^*$  \\
&   {\cite{adams}}& & & & & & 16$^0$ & 1.28 &
.76 & .78 & .92, .88$^*$ & .94, .90$^*$ \\
 & & & & & & & 14$^0$ & 1.29 &
.77 & .79 & .93, .89$^*$ & .95, .91$^*$ \\ \hline

 & & & & & & & 20$^0$ &
 -.26 & -.30 & -.31  & -.32, -.36$^*$ & -.34, -.38$^*$ \\
 $\Delta$ d & -.41  & -.33 & -.35 & -.37 & -.40 & -.41 & 
18$^0$ & -.27 &-.31 & -.32 & -.33, -.37$^*$ & -.35, -.39$^*$ \\
   &  {\cite{adams}} & & & & & & 16$^0$ &
  -.28 & -.32 &-.33 & -.34, -.38$^*$ & -.36. -.40$^*$ \\
   & & & & & & & 14$^0$ &
 -.29 & -.33 & -.34 & -.35, -.39$^*$ & -.37, -.41$^*$ \\ \hline
&&&&&&&&&&&& \\
$\Delta$ s &-.07  & 0 & -.1 & -.12 & -.02 & -.02 & &
0 & -.1 & -.12 & -.02, -.06$^*$ &-.02, -.06$^*$ \\  
&  {\cite{adams}} &&&&&&&&&&& \\ \hline
 & & & & & & & 20$^0$ &
1.52 & 1.04 & 1.07 & 1.22, 1.22$^*$ & 1.26, 1.26$^*$ \\
$G_A/G_V$ & 1.267  & 1.66 & 1.14 & 1.18 & 1.35 & 1.40 & 18$^0$ &
1.54 & 1.06 & 1.09 & 1.24, 1.24$^*$ & 1.28, 1.28$^*$  \\
 & {\cite{PDG}} & & & & & & 16$^0$ &         
1.56 & 1.08 & 1.11 & 1.26, 1.26$^*$ & 1.30, 1.30$^*$  \\
 & & & & & & & 14$^0$ &
1.58 & 1.10 & 1.13 & 1.28, 1.28$^*$ & 1.32, 1.32$^*$ \\   \hline

\end{tabular}
\end{center}}
* {\small Values after inclusion of the contribution from anomaly
{\cite{anomaly}}.}
\caption{The calculated values of spin polarization functions and $G_A/G_V$.}
\end{table}

In Table 1, we have presented the results of our calculations.
First of all, for $\chi$QM with SU(3) symmetry breaking as
well as configuration mixing, we have carried out a  $\chi^2$ fit to $\Delta u, ~\Delta d, ~
\Delta s, ~G_A/G_V$ and other related parameters
(details of which will be published elsewhere), however, in the fit 
we have taken $\phi \simeq$ 20$^0$, a value dictated by consideration of
neutron charge radius {\cite{yaouanc}}.
In the table we have also considered a few
more values of the mixing parameter $\phi$ in order to study the variation 
of spin distribution functions with  $\phi$.
The parameter $a$ is taken to be 0.1, as considered by other 
authors \cite{{cheng},{quigg},{johan},{song}}. 
The  symmetry breaking parameters obtained from  $\chi^2$ fit are 
$\alpha=.4$ and $\beta=.7$
for the data corresponding to recent E866 {\cite{E866}} as well as  
NMC {\cite{NMC}} data . The parameter $\zeta$ is constrained by the 
expressions $\zeta=-0.7-\beta/2$ ~and~ $\zeta=-\beta/2$
for the NMC and E866 experiments respectively, which essentially
represent the fitting of deviation from Gottfried sum rule {\cite{GSR}}.
Further, while presenting the results of  SU(3) symmetry 
breaking case without configuration mixing $(\phi=0^0)$, 
we have used the same values 
of parameters  $\alpha$ and  $\beta$, primarily to understand the role of
configuration mixing for this case. 
The SU(3) symmetry calculations based on Equations (13), (14) and (15) 
are obtained by taking $\alpha= \beta=1, \phi=20^0$ and  
$\alpha= \beta=1, \phi=0^0$ respectively for with and without 
configuration mixing. 
For the sake of completion, we have also presented the results of
CQM with and without configuration mixing. In this case the spin
polarization functions can  easily be found from 
equations (10), (11) and (12), for example,

\be
 \Delta u = cos^2 \phi[\frac{4}{3}] + sin^2 \phi[\frac{2}{3}],
\ee

\be
 \Delta d = cos^2 \phi[-\frac{1}{3}] + sin^2 \phi[\frac{1}{3}],
\ee

and

\be
 \Delta s = 0.
\ee

From Table 1, it is clear that $\chi$QM with SU(3) symmetry
breaking along with configuration mixing generated by one
gluon exchange forces is able to give an excellent
fit to the spin polarization data for symmetry breaking parameters
$\alpha=.4$ and $\beta=.7$.
In order to appreciate the role of
configuration mixing in affecting the fit, we first  compare the
results of CQM with those of CQM$_{gcm}$.
One observes that
configuration mixing corrects the result of the quantities
in the right direction but this is not to the desirable level.
Further, in order to understand the role of configuration mixing
and SU(3) symmetry with and without breaking  in $\chi$QM,
we can compare the results of $\chi$QM with SU(3)
symmetry to those of $\chi$QM$_{gcm}$ with SU(3)
symmetry. Curiously $\chi$QM$_{gcm}$ compares
unfavourably with  $\chi$QM
in case of the calculated quantities.
This indicates that configuration mixing alone is not enough to generate
 an appropriate fit in  $\chi$QM.
However when  $\chi$QM$_{gcm}$ is used with SU(3)  symmetry 
breaking then the results show uniform improvement over the corresponding
results of $\chi$QM with  SU(3) symmetry breaking.
It is interesting to note that the values of $\alpha$ and $\beta$ are in
agreement with other recent calculations {\cite{{cheng},{johan}, {song}}}.

After having seen that $\chi$QM$_{gcm}$  is able to
accomodate $G_A/G_V$, one turns to the question of compatibility
of $G_A/G_V$, $\mu_{p,n}$ and $<r_n^2>$. In this regard, we first
evaluate magnetic moments in the $\chi$QM$_{gcm}$. In this context,
it has been shown recently by Cheng and Li {\cite{cheng1}} that
 $\chi$QM, incorporating the symmetry
breaking effects, leads to formula which is similar to CQM, for
example,

\be
\mu(B)=(1+\kappa_{spin}+\kappa_{orbit}) \mu(B)_v,
\ee
where $(1+\kappa_{spin}+\kappa_{orbit})$ is the Cheng and Li
scale factor  and $\mu(B)_v$ is the magnetic moment in the CQM.
The Cheng and Li scale factor can be absorbed in the quark masses,
thus magnetic moments calculated in
$\chi$QM essentially reduce to using appropriate `quark mass scales' for
fitting the magnetic moments.

In Table 2, we have presented the magnetic moments with the 
wavefunctions expressed through  equation (6), 
in terms of $\mu_{56}$ and $\mu_{70}$,
\[ \mu(N) =cos^2 \phi<56|M|56> + sin^2 \phi <70|M|70> \] 
\[={cos}^2 \phi (\mu_{56}) +{sin}^2 \phi (\mu_{70}), \]
where $M$ is the magnetic moment operator. It is
evident from the table that for a good range of $\phi$ we can
fit the magnetic moments of the nucleons. As has been shown
 {\cite{mgupta1}}, this not only
reproduces nucleon magnetic moments but is also able to give a
good fit to other magnetic moments . Further, the magnetic moments
are independent of the sign of mixing angle $\phi$.

\begin{table}
\begin{center}
\begin{tabular}{|c|c|c|c|c|c|} \hline

 & $\phi$ & $\mu$(p) & $\mu$(n) & $R^2$ 
 & $<r_n^2>$  \\ 
&&&&   $(GeV^{-2})$ & $(GeV^{-2})$ \\  \hline
$\mu_{56}$ & - & $\frac{1}{3}$(4$\mu_u$ - $\mu_d$) & 
$\frac{1}{3}$(4$\mu_d$ - $\mu_u$) & -  & - \\ \hline
$\mu_{70}$ & & $\frac{1}{3}$(2$\mu_u$+$\mu_d$) & 
$\frac{1}{3}$(2$\mu_d+\mu_u$) & - & - \\ \hline
Expt value & -& 2.793 & -1.913 & - & 2.82 \\ \hline
Calculated  & -$20^0$ & 2.766 & -1.767 & 8 & 2.64 \\
 values                &    &      &       & 9 & 2.93 \\
                 &    &      &       & 10& 3.23 \\

             & -$18^0$ & 2.804 & -1.805 & 8 & 2.39 \\
                 &    &      &       & 9 & 2.65  \\
                 &    &      &       &10 & 2.91  \\

             & -$16^0$ & 2.841 & -1.843 & 10 & 2.61 \\
                 &    &      &       & 11 & 2.84 \\
                 &    &      &       & 12 & 3.07 \\

             & -$14^0$ & 2.882 & -1.882 & 12 & 2.71 \\
                 &    &      &       & 13 & 2.91 \\
                 &    &      &       & 14 & 3.11  \\ \hline

\end{tabular}
\caption{The magnetic moments and neutron charge radius for
different values of $\phi$ and $R^2$. The `effective masses' for 
the constituent quarks used here are
$m_u=m_d=0.313 ~GeV, ~m_s=0.513 ~GeV$ .}
\end{center}
\end{table}

After having realized that the $\chi$QM$_{gcm}$ can explain
$G_A/G_V$ and the nuclear magnetic moments, we have tried to
investigate whether it is able to reproduce  the neutron charge radius 
$<r^2_n>$. As it has been emphasized earlier that the Foldy term, 
reproducible within $\chi$QM, gets cancelled against a contribution to 
the Dirac form factor {\cite{Isgur}}, therefore to calculate the neutron 
charge radius we have effectively replaced
$ G_{E}^n (q^2)_{\chi QM} \Rightarrow   G_{E}^n (q^2)_{QM}. $
In the CQM similar calculations have been done \cite{mgupta2}. In order to 
emphasize its dependence on mixing angle $\phi$ we reproduce here some
of the essential details \cite{yaouanc}.

 The neutron charge radius is usually expressed in terms of the
 slope of the electric form factor $ G_{E}^n(q^2)$, the experimental
 value {\cite{kopecki}} of which is given as

 \be  (\frac{d G^n_E (q^2)}{d|q^2|})_{q^2=0} =
    0.47 \pm 0.01 ~GeV^{-2}.
 \ee

If we assign the nucleon to a pure 56 (with the spin expressed in
 terms of Pauli spinors ), the neutron electric form factor
 vanishes for all $q^2$.
Considering our complete wavefunction with the $56-70$ mixing the
neutron charge radius can be expressed as

 \be
 <r_n^2>= 6 (\frac{d G^n_E (q^2)}{d|q^2|})_{q^2=0} =
\sqrt{\frac{2}{3}} R^2 (-tan \phi),
 \ee
where $\phi$ is the mixing angle which is negative and $R^2$ is the
shape factor {\cite{{mgupta2},{yaouanc}} for the harmonic oscillator
wave function. We have given the calculated values of neutron charge
radius $<r^2_n> (=6b)$ as function of $\phi$ and $R^2$ in Table 2.
From the table one
finds that one is able to reproduce fairly well the experimental
value of $<r_n^2>$, {\it viz.}, 2.82~ GeV$^{-2}$ in the
$\chi$QM$_{gcm}$ by considering the same
values of parameter $\phi$ which reproduced the values of
$G_A/G_V$ and nuclear magnetic moments.

In conclusion, we would like to mention that we have calculated
 the spin polarization functions
$(\Delta u,~ \Delta d$ and $\Delta s)$ in the $\chi$QM 
with SU(3) symmetry breaking as well as one gluon generated 
configuration mixing for the NMC and the most recent E866 data. 
Besides reproducing  $G_A/G_V$ and magnetic moments, it can also  accomodate 
the neutron charge radius, thus resolving the compatibility problem  of 
these parameters which could neither be achieved in the CQM with
configuration mixing nor in the $\chi$QM with and without 
SU(3) symmetry breaking.

  \vskip .2cm
  {\bf ACKNOWLEDGMENTS}\\
M.G. would like to thank L.M. Sehgal for some useful discussions as
well as initiating his interest in the problem.
H.D. would like to thank CSIR, Govt. of India, for
 financial support and the Chairman,
 Department of Physics, for providing facilities to work in the department.

\end{document}